\documentclass[12pt,a4paper,reqno]{amsart}
\usepackage{latexsym}
\usepackage{amsmath}
\usepackage{amssymb}
\usepackage{graphicx}
\newcommand{\Lie}{{\mathcal{L}}}

\newcommand{\C}{\mathbb{C}}
\newcommand{\CP}{\mathbb{CP}}

\newcommand{\R}{\mathbb{R}}

\newcommand{\OO}{\mathcal O}

\newcommand{\mnote}[1]
{\protect{\stepcounter{mnotecount}}$^{\mbox{\footnotesize
$
\bullet$\themnotecount}}$ \marginpar{
\raggedright\tiny\em
$\!\!\!\!\!\!\,\bullet$\themnotecount: #1} }

\def\be{\begin{equation}}
\def\ee{\end{equation}}

\def\p{\partial}

\begin{document}
\date{January 15th, 2023}
\title{Quantum state reduction, and Newtonian twistor theory}
\author{Maciej Dunajski}
\address{Department of Applied Mathematics and Theoretical Physics\\ 
University of Cambridge\\ Wilberforce Road, Cambridge CB3 0WA\\ UK.}
\email{m.dunajski@damtp.cam.ac.uk}
\author{Roger Penrose}
\address{The Mathematical Institute, University of Oxford, Andrew Wiles Building, Woodstock Road, Oxford OX2 6GG, UK.  }
\email{rpenroad@gmail.com}
\begin{abstract}
We discuss the equivalence principle in quantum mechanics in the context of Newton--Cartan geometry, and non--relativistic twistor theory.
 \end{abstract}
\maketitle
\section{Introduction}
In a series of papers \cite{Pe96, Pe11} one of us (RP) argued  that to resolve the measurement paradox in quantum mechanics gravitational effects
have to be taken into account. The arguments do not rely on gravitational fields being strong, and in fact
only use Newtonian gravity. Newtonian gravity has recently been incorporated into the framework of twistor theory \cite{DG16}. This version
of twistor theory describes all Newton--Cartan space--times and is not constrained by the self--duality obstruction which has so far
prevented incorporating the full Lorentzian Einstein equations into the twistor framework.

Moreover, whatever it is that actually goes on physically when
the wave-function collapses, i.e. the reduction of the quantum state, or {\bf R}-process-this being taken to be an objective physical process-must have a curious
`retro-active' aspect to it if taken to be a `classically real' physical process (see \cite{P22}). Such ontological puzzles do not present difficulties in the Newtonian limit, so it makes sense, at our present level of understandings to concentrate on the Newtonian situation, where such issue of the precise timing of the {\bf R}-process can be evaded.

Quantum
theory is inherently non--local, and this non--locality is likely to prevail in any  modifications resulting
from incorporating gravitational fields. Newtonian twistor theory is also non--local: space--time points correspond to extended and global objects (rational curves) in the twistor space.
 It is hoped, and has been suggested in \cite{ADM17}, that combining  quantum non--locality with  Newtonian twistor non--locality can shed light on 
the role of gravity in quantum state reduction. In this paper we aim to make a connection between the two theories.

 Our focus  will be the principle of equivalence applied to a quantum particle in a gravitational field,
which results in a time--dependent phase difference between  wave functions in different reference frames.
The Schr\"odinger equation for a particle moving in a  uniform field given by ${\bf g}=(0, 0, g)$ where $g$ is  a constant is
\be
\label{schr_1}
i\hbar \frac{\p \psi}{\p t}=-\frac{\hbar^2}{2m}\frac{\p^2 \psi}{\p z^2}-mgz \psi.
\ee
This is the Newtonian reference frame, and the corresponding $\psi$ has in \cite{Pe11} been called
the Newtonian wave function. The same particle can be considered in the freely falling 
frame - called the Einstein frame  - related to the Newtonian frame by
the transformation
\be
\label{transformz}
z=Z+\frac{1}2 T^2g, \quad t=T.
\ee
The corresponding Schr\"odinger equation is
\be
\label{schr_2}
i\hbar \frac{\p \Psi}{\p T}=-\frac{\hbar^2}{2m}\frac{\p^2 \Psi}{\p Z^2},
\ee
and using the chain rule one finds
\be
\label{phase}
\psi(z, t)=\Lambda(z, t) \Psi(Z, T), \quad \mbox{where}\quad 
\Lambda=e^{-\frac{im}{\hbar}\Big(\frac{t^3g^2}{6}-tgz\Big)}.
\ee
The non--linear time--dependence of the phase factor 
leads to a discrepancy between the notions of positive  and negative frequencies
in Newtonian and Einsteinian frames. Extrapolating these ideas to quantum field theory would lead to two different vacua in two different Hilbert spaces which poses a problem for the standard QFT framework.

In \S\ref{section2} we shall argue that the  cubic time dependence in the phase ambiguity (\ref{phase})
is also present in non--uniform gravitational fields given by a general Newtonian potential
$V:\R^3\rightarrow \R$. We shall employ the Eisenhart lift, where the Schr\"odinger equation
arises as a null reduction of the wave equation on the curved plane-wave background in $4+1$ dimensions. The diffeomorphism between the Newtonian frame, and the Einstein frame in Riemann normal coordinates, where the Newtonian connection vanishes
at a point is generated by a translation along the parallel null direction, which in turn gives rise
to a phase difference
\[
\Lambda({\bf x}, t)=\exp{{\Big(-\frac{im}{\hbar} {\Big(\frac{t^3}{6m^2}|{\bf \gamma}|^2+\frac{t}{m}{\bf x}\cdot{\bf \gamma}\Big)\Big)}}}, \quad\mbox{where}\quad
{\bf \gamma}=\nabla V|_{{\bf x}={\bf 0}}.
\]
The vector $\gamma$ contains the non--vanishing components of the Newtonian connection, and in \S\ref{section2} we shall also discuss the interplay between
quantum phase and the curvature of this connection.

In \S\ref{section3} we shall focus on the uniform field, where stationary states in the Newtonian frame
correspond to solutions of the Airy equation. This equation admits a twistor description \cite{CD}
in terms of cohomology classes invariant under a $3$--dimensional abelian subgroup of the full conformal group of Minkowski space--time $M$. We shall show that the action of this subgroup preserves a light--ray in 
$M$, and a point on it. The twistor function representing the Airy cohomology class is the exponential of a cubic in a twistor variable parametrising $\alpha$--planes through a point in $M$. This cubic dependence
formally resembles the time dependence  of the phase ambiguity (\ref{phase}). In \S\ref{section31} we shall explain this by the twistor--Fourier transform to to the momentum space.

In \S\ref{section4} we shall review non--relativistic twistor theory \cite{DG16}, with an emphasis on the inverse twistor function for the uniform gravitational field. Using this twistor function
to deform the patching relations of the non--relativistic twistor space leads to a holomorphic instability:
the normal bundle of  non--relativistic twistor curves jumps from $\OO\oplus\OO(2)$ to $\OO(1)\oplus\OO(1)$
which is the relativistic normal bundle of curves in the non--linear graviton construction \cite{Pen_3}.
We shall use this mechanism to suggest a proposal for a twistorial collapse of the wave function: while
the space--time geometry may bifurcate during the collapse, the twistor space remains a complex three--fold, but one four--dimensional family of rational curves needs replacing by another such family.
\subsection*{Acknowledgments} 
The work of MD has been partially supported by STFC grants ST/P000681/1, and  ST/T000694/1. We thank Gary Gibbons, Peter Horv\'athy, and  Paul Tod for useful discussions, and Adam Dunajski for his help in producing Figure 1.
\section{The phase shift for non--uniform gravitational fields.}
\label{section2}

The phase ambiguity (\ref{phase}) is of little concern as either the Einstenian, or the Newtonian
wave function can be chosen to describe the particle, and used consistently to model its  quantum evolution.
The problem arises if the gravitational field itself is involved in  quantum superposition. An example
considered in \cite{Pe11} is a massive object placed in a superposition in two locations ${\bf 1}$ 
and ${\bf 2}$, each with its own uniform gravitational field ${\bf g}_1$ and ${\bf g}_2$. An Einsteinian wave function can still be considered at a point $p$ and (as the overall gravitational field of the earth cancels out) there is a phase difference of the form (\ref{phase}) with $g^2$ replaced by $|{\bf g}_1-{\bf g}_2|^2$ and $gz$ replaced by $({\bf g}_1-{\bf g}_2)\cdot {\bf x}$.

 In this section we shall show that the  cubic time dependence in the 
phase ambiguity is not an artifact of adopting a uniform gravitational field, and is present
for general potentials in the Schr\"odinger equation.

We shall use a geometric setup, where the time--dependent Schr\"odinger equation arises 
from a null Kaluza--Klein reduction\footnote{Such reduction have been studied in various context
following \cite{duval} and \cite{duval2}. A closer examination reveals
that they go back to Eisenhart \cite{eisenhart}.}.
Consider a plane--wave metric in $4+1$ dimensional manifold ${\mathcal M}^5$ with local coordinates
$x^{\mu}=(u, t, x^i)$ with a null isometry  which we 
chose to be generated by a null Killing vector $K=\p/\p u$ and the line element 
\be
\label{eisenhart}
G=2 dudt +2 \frac{V({\bf x}, t)}{m} dt^2- d {\bf x}\cdot d{\bf x}.
\ee
where $m$ is a constant with the dimension of mass.
Let  $\phi(u, t, {\bf x})$ be a complex--valued function  which
satisfies the wave equation with respect to the metric $G$, and is invariant under 
$K$
\[
\Box_G\phi=0, \quad K(\phi)=-\frac{im\phi}{\hbar}.
\]
Solving the invariance
condition as 
\be
\label{wave_sh}
\phi(u, t, {\bf x})=e^{-\frac{imu}{\hbar}}\psi(t, {\bf x})
\ee
and substituting this to the wave
equation yields the Schr\"odinger equation\footnote{
According to the proposal of
\cite{Pe96}, a superposition of two states corresponding to lumps of matter should spontaneously collapse to stationary states of the 
Schr\"odinger--Newton
system  for a quantum particle moving in its own gravitational potential \cite{moroz} 
\be
\label{schrodinger_n}
i\hbar \frac{\p \psi}{\p t}=-\frac{\hbar^2}{2m}\Delta \psi+V\psi, \quad
\Delta V=4\pi Gm |\psi|^2.
\ee
The anzats (\ref{wave_sh}) gives the first equation in (\ref{schrodinger_n}).
Imposing the Einstein equation with the null dust energy--momentum tensor 
\[
R_{\mu\nu}=4\pi G |\phi|^2 \theta_\mu \theta_\nu, \quad \mbox{where}\quad \theta=dt
\]
yields the second equation in (\ref{schrodinger_n}).
 Thus the Schr\"odinger--Newton system (\ref{schrodinger_n})
arises from a coupled system consisting of the Einstein equation with a null energy--momentum tensor, and the 
invariant wave equation for the $(tt)$ component of this tensor.
}
\be
\label{schrodinger}
i\hbar \frac{\p \psi}{\p t}=-\frac{\hbar^2}{2m}\Delta \psi+V\psi.
\ee
The null geodesics in $({\mathcal M}^5, G)$ project to paths on 
the four--dimensional space of orbits $N$ of the null isometry $\p/\p u$ in ${\mathcal M}^5$.
These paths satisfy the Newton equation
\[
m\frac{d^2 {\bf x}}{d t^2}=-\nabla V,
\]
and are unparametrised geodesics of the Newton--Cartan connection with 
the non--vanishing connection components
\be
\label{new_connection}
\gamma_{tt}^i=\frac{1}{m}\delta^{ij}\frac{\p V}{\p x^j}, \quad i, j, \dots = 1, 2, 3.
\ee
The curvature tensor of this connection is given by
\be
\label{new_cur}
{r^i}_{tjt}=\frac{1}{m}\frac{\p {\gamma^i}_{tt}}{\p x^j},
\ee
with all other components vanishing identically.
\subsection{Uniform gravitational field field}
If the gravitational field is uniform 
\be
\label{lin_potential}
V=-m{\bf g}\cdot{\bf x}
\ee
then the connection components (\ref{new_connection}) are constant, and equal to
$-g^i$, and
can be set to zero locally by going to the accelerating frame
$T=t, {\bf X}={\bf x}-(1/2){\bf g} t^2$ which is what we did in 
(\ref{transformz}). Geometrically this corresponds to the fact that the
curvature of the Newtonian connection (\ref{new_cur}) vanishes for constant ${\bf g}$. 

The corresponding Eisenhart metric (\ref{eisenhart}) is also flat. 
The calculation below shows that a transformation to flat coordinates involves a non--constant
translation of the $u$--coordinate which, at the level of (\ref{wave_sh}), corresponds to a non--constant phase shift of $\psi$.
The metric admits five-dimensional Abelian group isometries generated by the Killing 
vectors $K_u, K_t, K_j$, where $j=1, 2, 3$ 
and 
\begin{eqnarray*}
K_u&=&\frac{\p}{\p u},\\
\quad K_{t}&=& \frac{\p}{\p t}+tg^j\frac{\p}{\p x^j}+\Big(\frac{1}{2}|{\bf g}|^2t^2+{\bf g}\cdot{\bf x} \Big)\frac{\p}{\p u},\\
\quad K_j&=&-\frac{\p}{\p x^j}-g_j t\frac{\p}{\p u}, \quad j=1, 2, 3.
\end{eqnarray*}
These Killing vectors correspond to translations in the flat coordinates, and their metric--dual one--forms are necessarily gradients:
\[
G(K_u, \cdot)=dT, \quad G(K_t, \cdot)=dU, \quad G(K_j, \cdot)=dX^j,
\]
where
\be
\label{coords1}
T=t, \quad U=u-t{\bf g}\cdot {\bf x}+\frac{1}{6}|{\bf g}|^2t^3, \quad {\bf X}={\bf x}-\frac{1}{2}{\bf g}t^2
\ee
and using $(U, T, X^i)$ as coordinates yields
\[
G=2dUdT-d {\bf X}^2.
\]
Performing the inverse coordinate transformation at the level of wave functions (\ref{wave_sh}) from the flat coordinates,
where the Schr\"odinger equation has zero potential, to the Newtonian coordinates $(u, t, {\bf x})$ where the potential
is (\ref{lin_potential}) gives
\begin{eqnarray}
\label{phase_normal}
\Phi(U, T, {\bf X})&=&e^{-\frac{imU}{\hbar}}\Psi({\bf X}, T)\nonumber\\
&=&e^{-\frac{imu}{\hbar}}\Lambda({\bf x}, t) \Psi({\bf X}, T)\\
&=&e^{-\frac{imu}{\hbar}}\psi({\bf x}, t), \quad\mbox{where}\quad
\Lambda({\bf x}, t)=e^{-\frac{im}{\hbar}\Big(\frac{1}{6}|{\bf g}|^2t^3-t{\bf g}\cdot{\bf x} \Big)}.\nonumber
\end{eqnarray}
This is in agreement with the phase shift  $(\ref{phase})$ where 
\[
\psi({\bf x}, t)=\Lambda({\bf x}, t)\Psi\Big({\bf x}-\frac{1}{2}{\bf g}t^2, t\Big).
\]
\subsection{Non--uniform gravitational field}
In the 
general non--uniform, but weak, gravitational field we can chose the 
normal coordinates centred at some point $p$ in $N$, such that ${\gamma^i}_{tt}(p)=0$.
The transformation to these coordinates can be performed at the level of the 
Schr\"odinger equation with the potential $V$. The equivalence principle
does not allow a transformation to a free--falling Einsteinian frame
(there will be tidal effects caused by non--zero 
curvature which can not be transformed away, as the curvature is a tensor), 
but the cubic time dependence in the phase factor 
(\ref{phase}) turns out to be universal.
The free--falling Einstein frame  is replaced by the Riemann normal coordinates
$X^{\alpha}=(U, T, X^i)$ such that
\be
\label{riemann_cor}
G=2dUdT-{d\bf X}^2+\frac{1}{3}R_{\mu\alpha\beta\nu}(0)X^{\alpha}X^{\beta} dX^{\mu} dX^{\nu}+O(|X|^3),
\ee
where ${R^{\gamma}}_{\mu\alpha\beta}$ is the Riemann curvature of $G$, and the index has been lowered with $G$.
The Riemann coordinates are given in terms of the original Newtonian cordinates
$x^{\mu}=(u, t, x^i)$ by
\be
\label{riemann_2}
X^{\mu}=x^{\mu}+\frac{1}{2}{\Gamma^{\mu}}_{\alpha\beta}(0)x^{\alpha}x^{\beta}
+\frac{1}{6}({\Gamma^{\mu}}_{\alpha\gamma}(0){\Gamma^{\gamma}}_{\beta\delta}(0)
+\p_\delta{\Gamma^{\mu}}_{\alpha\beta}(0))x^{\alpha}x^{\beta}x^{\delta}+O(|x|^4)
\ee
where ${\Gamma^{\mu}}_{\alpha\beta}$ are the Christoffel symbols of $G$ in coordinates $x^{\mu}$.
The quantities
\be
\label{gammar}
\gamma^i=\delta^{ik}\p_k V|_{{\bf x}={\bf 0}}, \quad    {r^i}_j=\delta^{ik}\p_j\p_k V|_{{\bf x}={\bf 0}}
\ee
encode the Newtonian connection, and its curvature at a point $p$ with coordinates $(0, {\bf 0})$.
Computing (\ref{riemann_2}) for the Eisenhart metric (\ref{eisenhart}) with $V=V({\bf x}, t)$ gives
\begin{eqnarray*}
T&=&t,\\
X^i&=&x^i+\frac{t^2}{6m}(x^j\p_j\p_k V+3\p_k V+t\p_k\dot{V})\delta^{ik},\\
U&=&u+\frac{t^3}{6m^2}|\nabla V|^2+\frac{t}{3m}x^ix^j\p_i\p_j V+\frac{t}{m}x^i\p_i V\\
&& +\frac{t^3}{6m}\ddot{V}+\frac{t^2}{2m}(\dot{V}+x^i\p_i\dot{V}),
\end{eqnarray*}
where the derivatives of $V$ in these formulae are evaluated at $(0, {\bf 0})$.
Assuming that $V$ does not explicitly depend on $t$, and using vector (\ref{gammar}) notation on $\R^3$ this becomes
\be
\label{vec_from}
U=u+\frac{t^3}{6m^2}|{\bf \gamma}|^2+\frac{t}{3m}{\bf x}^T r \;{\bf x}+\frac{t}{m}{\bf x}\cdot {\bf \gamma}, \quad
{\bf X}={\bf x}+\frac{t^2}{6m} r{\bf x}+\frac{t^2}{2m}{\bf \gamma}.
\ee
This transformation puts the Eisenhart metric in the Riemann coordinates form (\ref{riemann_cor}) but  it does not preserve the form  of the metric, as
it brings up the curvature terms proportional to
$r_{ij}TX^i dX^j dT$ and $r_{ij}T^2dX^idX^j$.
Therefore we shall truncate (\ref{vec_from}) to
\be
\label{truncation}
U=u+\frac{t^3}{6m^2}|{\bf \gamma}|^2+\frac{t}{m}{\bf x}\cdot {\bf \gamma}, \quad
{\bf X}={\bf x}+\frac{t^2}{2m}{\bf \gamma}.
\ee
In particular the $({\bf x}, t)$--dependent shift in the null direction $u$ results in the phase-shift of the 
Schr\"odinger wave function as
\[
e^{-\frac{mU}{\hbar}}\Psi({\bf X}, T)= e^{-\frac{mu}{\hbar}}\psi({\bf x}, t),
\]
or
\[
\psi({\bf x}, t)=\Lambda({\bf x}, t)\Psi({\bf x}+\frac{t^2}{2m}{\bf \gamma}, t)
\]
with the  phase factor
\[
\Lambda({\bf x}, t)=e^{-\frac{im}{\hbar} {\Big(\frac{t^3}{6m^2}|{\bf \gamma}|^2+\frac{t}{m}{\bf x}\cdot{\bf \gamma}\Big)}}, \quad
{\bf \gamma}=\nabla V|_{{\bf x}={\bf 0}}.
\]
The non--vanishing connection and curvature of the Eisenhart metric 
are
\[
{\Gamma^i}_{tt}={\gamma^i}_{tt}, \quad {\Gamma^u}_{tj}=\frac{1}{m}\frac{\p V}{\p x^j},\quad
 {R^i}_{tjt}={r^i}_{tjt}, \quad  {R^u}_{itj}=\frac{1}{m}\frac{\p^2 V}{\p x^i \p x^j},
\]
where ${\gamma^i}_{tt}=m^{-1}\delta^{ij}\p_j V$ and ${r^i}_{tjt}=\p_j {\gamma^i}_{tt}$ are the non--vanishing components of the corresponding Newtonian connection and curvature on $N$. Therefore if $G$ is flat, then so is the Newtonian connection.  Moreover if $(T, U, X^i)$ is a Riemann normal coordinate system centered
at a point $P\in {\mathcal M}^5$, then $(T, X^i)$ are normal coordinates on the Newtonian space--time
centered at $\pi(P)$, where $\pi$ is the projection from the plane--wave space--time to the Newtonian space--time. The truncation (\ref{truncation}) agrees with the Riemann coordinates up to the lowest order term, and still ensures that the Newtonian connection vanishes at $p$. The corresponding frame of reference is a curved analogue of the free--falling Einstein frame.
 The Einstenian and Newtonian wave functions  differ by a time--dependent phase factor
which -- if the equivalence principle is taken seriously -- leads to ambiguities in a definition of positive and negative frequency.

\section{Twistor theory of the Airy equation}
\label{section3} 
The stationary states of a quantum particle in the linear potential (\ref{schr_1})
are $\psi(z, t)={\mathcal T}(t) W(z)$, where
\[
{\mathcal T}=e^{-itE/\hbar}\quad\mbox{and}\quad
\frac{d^2W}{dz^2}+\Big(\frac{2m^2g}{\hbar^2}z+\frac{2mE}{\hbar^2} \Big)W=0, 
\]
where $E$ is a constant.
Setting
$\zeta={(2m^2g/\hbar^2)}^{1/3}(z+E/(gm))$ in the time--independent Schr\"odinger equation 
we recognise the Airy equation 
\be
\label{airy}
W''+\zeta W=0.
\ee

The Airy equation arises as a symmetry reduction of the wave equation on the 3+1--dimensional Minkowski space.  This leads to a twistor description \cite{CD} of (\ref{airy}). The details are as follows.
The wave equation
\be
\label{waveeq}
\frac{1}{c^2}\frac{\p^2 \phi}{\p t^2}-\frac{\p^2 \phi}{\p x^2}-
\frac{\p^2 \phi}{\p y^2}-\frac{\p^2 \phi}{\p z^2}=0.
\ee
on the Minkowski $M$ space with the metric
\[
\eta=c^2dt^2-dx^2-dy^2-dz^2
\]
is solvable in terms of the twistor contour integral formula \cite{Pen_2}
\be
\label{contour}
\phi(x,y,z,t)=\oint_{\Gamma\subset \CP^1}
f({{(z+ct)}}+{{(x+i y)}}\lambda,{{(x-i y)}} 
-{{(z-ct)}}\lambda, \lambda)d\lambda,
\ee
where $\Gamma\subset \CP^1$ is a closed contour in the twistor line $L_p\cong \CP^1$
corresponding to a point $p\in M$ with coordinates $(ct, x, y, z)$. In \cite{CD}
it was pointed out that solutions to (\ref{waveeq}) invariant under a certain three--dimensional Abelian subgroup $H$ of the conformal group of $(M, \eta)$ are
characterised by the Airy functions. The computations in \cite{CD} have been carried
in the complexified setup, and here we summarise the results imposing 
Lorentzian reality conditions. The subgroup $H\subset SU(2, 2)$ is generated by a null
translation, a non--null translation and a combination of a rotation, a Lorentz boost and another non--null translation which we chose to be
\begin{eqnarray*}
V_1&=&\frac{1}{2}\Big(\frac{\p}{\p z}+\frac{1}{c}\frac{\p}{\p t}\Big),\\
V_2&=&\frac{\p}{\p y},\\ 
V_3&=&\frac{1}{2}\Big(\frac{\p}{\p z}-\frac{1}{c}\frac{\p}{\p t}\Big)
+
\Big(\frac{x}{c}\frac{\p}{\p t}-ct\frac{\p}{\p x}\Big)+\Big(x\frac{\p}{\p z}-z\frac{\p}{\p x}\Big).
\end{eqnarray*} 
Now consider a function $\phi:M\rightarrow \R$ such that
\[
V_1(\phi)=\phi, \quad V_2(\phi)=0, \quad V_3(\phi)=0.
\]
We find that the general solution of this invariance condition is
\[
\phi=\exp{\Big((ct+z)+2(ct-z)x+\frac{2}{3}(ct-z)^3 \Big)}W(\zeta)
\]
where 
\[
\quad \zeta=-2x-(ct-z)^2.
\]
Imposing the  wave equation (\ref{waveeq}) on $\phi$ gives the Airy equation (\ref{airy})

We now move to the contour integral formula (\ref{contour}). Any generator of the conformal group of $M$ gives rise to a holomorphic vector field
on the twistor space $PT=\CP^3-\CP^1$. Using the twistor incidence relations we can find the generators of $H\subset SL(4, \C)$ as vector fields on $PT$. Let
$\Omega=fd\lambda$ be the integrand in (\ref{contour}), where the twistor function $f$ represents a cohomology class $H^{1}(L_p, \OO(-2))$.

Imposing the invariance condition 
\[
\Lie_{V_1}\Omega=\Omega, \quad \Lie_{V_2}\Omega=0, \quad \Lie_{V_3}\Omega=0,
\]
reduces (\ref{contour}) to the integral formula
\be
\label{airy_twistor}
W(\zeta)=\oint_{\Gamma\subset \CP^1} \exp{\Big(\frac{1}{3}\lambda^3+\lambda \zeta \Big)}d\lambda
\ee
for the Airy equation. Two linearly independent solutions to the Airy equation result 
from different choices of contours which pass through the essential singularity of the twistor function at $\lambda=\infty$. 

Integrating the vector fields $(V_1, V_2, V_3)$ to one--parameter subgroups on $PT$ we
find that the resulting subgroup $H$ of $SL(4, \C)$ is generated by matrices of the form
\[
\left (
\begin{array}{cccc}
1 & p &q &r\\
0 &1& p & q\\
0 & 0& 1& p\\
0 & 0& 0& 1
\end{array}
\right ).
\]
The projective action of $H$ can now be seen to preserve the flag
\[
\alpha\subset L \subset \beta
\]
on $PT$, where $\beta$ is a plane of the form $[*, *, *, 0]$, $L$ is a line 
$[*, *, 0, 0]$ and $\alpha$ is point $[*, 0, 0, 0]$. In the Minkowski space
$\beta$ corresponds to a $\beta$--plane, the line $L$ corresponds to a point 
and the point $\alpha$ corresponds to an $\alpha$--plane \cite{PR}. The intersection
of  $\alpha$ and  $\beta$ planes is (assuming these planes intersect in a non--empty set) in a null ray. The twistor incidence then implies that this null ray contains the point in $M$ corresponding to the line $L$.  Therefore the subgroup $H$ of the conformal group on $M$ preserves  a light ray and a point on it.
\subsection{Twistor--Fourier transform}
\label{section31}
The general solution  of the Airy equation is given by the twistor contour integral formula (\ref{airy_twistor}). 
The twistor function $e^{\lambda^3/3+\lambda \zeta}$ is formally similar
to the phase factor (\ref{phase}), except the exponential in the latter is cubic in time, and the exponential in the former is cubic in the twistor variable $\lambda$. The discussion below reconciles this.
\vskip5pt
In the Newtonian frame the Fourier transform of the Schr\"odinger equation (\ref{schr_1}) leads to the general
solution 
\begin{eqnarray}
\label{fourier}
\psi&=& \int_{\R} k\Big(p-\frac{mg}{\hbar}t\Big) e^{i\Big(-\frac{\hbar^2}{6m^2g}p^3+pz\Big)} dp\\
   &=& \Lambda(z, t) \int_\R k(p) e^{i\Big(p(z-\frac{1}{2}gt^2)-\frac{\hbar}{2m}p^2t -\frac{\hbar^2}{6m^2g}p^3\Big)} dp,\nonumber 
\end{eqnarray}
where
\[
\Lambda= e^{\frac{im}{\hbar}\Big(-\frac{t^3 g^2}{6}+gtz\Big)}
\]
where $k$ is an arbitrary function of one variable $p-mgt/\hbar$. 
To get from the first line to the second we have made a change of variables $p\rightarrow p+mgt/\hbar$, and took all the $p$--independent terms outside the integral. Now consider the initial configuration $\psi(z, 0)=\psi_0(z)$ with the Fourier
transform $\widehat{\psi}_0(p)$, so that 
\[
\psi_0(z)=\int_\R \widehat{\psi}_0(p) e^{ipz} dp.
\]
Comparing this  with (\ref{fourier}) evaluated at $t=0$ gives
\[
k(p)=e^{ip^3\hbar^2/(6m^2g)}\widehat{\psi}_0(p).
\] Substituting this into (\ref{fourier}),
and introducing coordinates $(Z, T)$ by (\ref{transformz})
gives
\be
\label{fourier2}
\psi(z, t)=\Lambda(z, t) \int_{\R}\widehat{\psi}_0(p) e^{i(pZ-\frac{\hbar^2}{2m}p^2T)}dp.
\ee
To recognise the integral term in (\ref{fourier2}) we note that the free Schr\"odinger equation (\ref{schr_2}) in the Einsteinan frame of reference
has the general Fourier solution 
\be
\label{fourier3}
\Psi(Z, T)=\int_R\widehat{\Psi}_0(p) e^{i(pZ-\frac{\hbar^2}{2m}p^2T)}dp.
\ee
Therefore, assuming that $\psi$ and $\Psi$ agree at $t=T=0$,  and comparing
(\ref{fourier2}) and (\ref{fourier3}) gives 
\[
\psi(z, t)=\Lambda(z, t)\Psi(Z, T)
\]
in agreement with formulae (\ref{phase}). 

We finish off with a comment about bound states (although it may not be relevant, 
or  even applicable to the  argument
in \cite{Pe11}). The Einsteinian wave function $\Psi$ is a plane wave, so it is not normalisable. Let us consider the Newtonian wave function
$\psi$ instead. It is a linear combination of the two Airy functions $Ai$ and $Bi$. The $Bi$ function diverges for large $\zeta$, so we shall
consider $\psi$ to be a multiple of $Ai$. Imposing the boundary condition $\psi(0, t)=0$ gives $Ai((2/(\hbar^2g^2m))^{1/3}E)=0$. The set of zeroes
of the Airy function $Ai$ is discrete, and this leads to a discrete set of energy levels. The linear potential $V=z$ for $z>0$ and $V=\infty$ for $z<0$  (known as the quantum
bouncer) with this boundary condition has been used to model ultracold  neutrons in the Earth gravitational field \cite{nish}.
\section{Inverse Newtonian twistor function}
\label{section4}
The Newtonian twistor space \cite{DG16, Gundry} consists of a complex three--manifold
$PT_{\infty}={\mathcal O}\oplus {\mathcal O (2)}$ thought of as the total space of rank--two holomorphic
vector bundle over $\CP^1$ (here $\OO(n)\rightarrow \CP^1$ denotes the holomorphic line bundle whose sections
are homogeneous polynomials on $\C^2$ of degree $n$) together
with a cohomology class 
\be
\label{fminus4}
f_{(-4)}\subset H^1(PT_{\infty}, {\mathcal O}(-4)).
\ee
The points of the corresponding Newton--Cartan space--time $N$ are holomorphic sections
of $PT_{\infty}\rightarrow \CP^1$. The global twistor function $T$ coordinatising
the ${\mathcal O}$ factor gives rise to a closed one--form (the {\em clock}) $\theta$ on $N$ which we chose to be $\theta=dt$. This gives a fibration of $N$ over the real--line (the time axis) which in turn gives the simultaneity of events in $N$.
There is a degenerate contravariant metric on $N$ defined by declaring a vector
in $T_pN$ to be null iff the corresponding section of the normal--bundle of $L_p\subset PT_{\infty}$ has a zero. Thus  null vectors are tangent to the fibers of constant $t$ in $N$.
If the homogeneous coordinates on $PT_{\infty}$ are $(T, Q, \pi_{A'})\sim(T, s^2 Q, s\pi_{A'})$ where $s\in \C^*$, then the twistor
lines are
\be
\label{section}
T=t, \quad Q=x^{A'B'}\pi_{A'}\pi_{B'},
\ee
where $x^{A'B'}=x^{(A'B')}$ are three coordinates on the spatial fibres on $N$.
The condition that ${\delta Q}=0$ has a repeated root leads to the flat metric 
\[
h=(\p/\p x)^2+(\p/\p y)^2+(\p/\p z)^2=\delta^{ij}\frac{\p}{\p x^i}\otimes\frac{\p}{\p x^j}.
\]
 Finally to recover the Newtonian connection
we consider the group $H^0(L_p, {\mathcal N}\otimes (\odot^2 {\mathcal N}))$,
where $L_p$ is the twistor line corresponding to a point $p\in N$, and \[
{\mathcal N}={\mathcal O}\oplus {\mathcal O (2)}\]
 is the normal bundle to $L_p$. A decomposition of this group contains a factor $H^0(L_p, {\mathcal O}(2))$, which by the Serre duality
is isomorphic to $H^1(L_p, {\mathcal O}(-4))$, and $f_{(-4)}$ is an element of 
this\footnote{This is discussed in detail in
\cite{DG16}, where it is also shown how $PT_{\infty}$ arises form a limiting procedure
of a one--parameter of nonlinear graviton twistor spaces. In this limit
the normal bundle of of the twistor curves jumps from $\mathcal{O}(1)\oplus \mathcal{O}(1)$ to
$\mathcal{O}\oplus \mathcal{O}(2)$. Cohomology classes defining these nonlinear graviton twistor spaces
as complex manifolds give rise, in the limit,  to the cohomology class $f_{(-4)}$}. 
It gives rise to a spin 1 field on $N$ given by
\[
g_{A'B'}=\oint_{\Gamma\subset \CP^1} f_{(-4)} \pi_{A'}\pi_{B'} \pi \cdot d\pi,
\]
and finally to the Newton--Cartan connection with the only non--vanishing components
\[
\gamma^{i}_{tt}=g^{i}=\delta^{ij}\frac{\p V}{\p x^j}, 
\]
where we identify a three--dimensional vector index $i$ with a pair
of symmetric spinor indices using the isomorphism
$\C^3=\mbox{Sym}^{2}(\C^2)$.
The function $V$ on $N$ is the scalar potential 
for 
${g}_{A'B'}$ given by
\[
V=\oint_{\Gamma\subset \CP^1} f_{(-2)} \pi \cdot d\pi, \quad
\mbox{where}\quad f_{(-4)}=\frac{\p f_{(-2)}}{\p Q}.
\]
This is a particular case of the twistor integral formula (\ref{contour}), and
we find that $V$ is harmonic with respect to the inverse metric $h$ on the three--dimensional constant time fibres of $N$. 

Specifying to the uniform gravitational field ${\bf g}=(0, 0, g)$ or $V=gz$,
we find the inverse twistor function
\[
f_{(-2)}=\frac{gQ}{(\pi\cdot o)^2(\pi\cdot \iota)^2},
\]
where $(o, \iota)$ is a pair of constant spinors. This can be verified by
restricting $f_{(-2)}$ to the twistor line (\ref{section}), and computing a 
residue. The gravitational field corresponds to the cohomology class
which splits as a coboundary, and thus represents a trivial cohomology class.
This is a manifestation of the Newtonian equivalence principle: the constant
gravitational field can be eliminated by moving to the accelerating frame.
\subsection{The wave function collapse}
A puzzling feature of the  space-time picture of the wave function collapse (called the {\bf R}--process in \cite{Pe96, Pe11}) is
the lack of knowledge as to when the state reduction takes place. 
It is reasonable to assume that it
happens in between two  times $t_0$, and $t_1$, and
introduces enough curvature 
that the $\OO+\OO(2)$ Newtonian twistor space is 
deformed\footnote{It has been noted long time ago \cite{tod_1} that holomorphic jumping lines in the twistor space
correspond to discontinous changes  in space--time geometry.}. While this may correspond to a discontinuous
jump in the space--time structure we propose that the twistor space survives the reduction 
(see also \cite{PRTN}), but
the 4-parameter family of curves with $\OO+\OO(2)$ normal bundle disappear, and needs
to be replaced by a new family according to the following mechanism
\begin{enumerate}
\item
 The cohomology class $f\in H^1 (PT_{\infty}, \OO)$ is used to deform the patching relation for the 
$T$ direction on $PT_{\infty}$. This cohomology class 
 is related to $f_{(-4)}$ in (\ref{fminus4}) by
\[
f_{(-4)}=\frac{\p^2 f}{\p Q^2},
\]
and is non-zero only between $T_0$ and $T_1$
on the twistor space, which correspond to times $t_0, t_1$. Therefore it is smooth, but not
analytic\footnote{This part of the proposal goes beyond the holomorphic realm of twistor theory. This may be justified at
physical grounds as, unlike the unitary Schro\"dinger evolution, the wave function collapse is not a holomorphic process.
More work is needed to justify that the essential features of the Kodaira deformation theory \cite{kodaira}
still apply.}. 
\item
Outside the $[T_0, T_1]$ interval, there exist a global twistor function $T$,
and the twistor curves are confined to constant $T$ slices (each such slice is
the total space of the $\OO(2)$ bundle over $\CP^1$). At $T_0$
we cover a line by two open sets (called $U$ and $\widetilde{U}$ on Figure 1), and deform
\[
\tilde{T}=T+\frac{1}{c}f(Q, T),
\]
where the deformation parameter is proportional to the inverse of the speed of light.
This makes the normal bundle ${\mathcal N}$ of the curves jump to $\OO(1)+\OO(1)$, and illustrates the general phenomenon that if
$H^1(\CP^1, \mbox{End}({\mathcal N}))\neq 0$, then general Kodaira deformations do not preserve
the type of the normal bundle. In our case this cohomology group equals $\C$.
 The blue $\OO+\OO(2)$ curves corresponding
to the Newtonian space time before the
measurement disappear. At $T_1$ another non-relativistic limit is taken, and
the new $\OO+\OO(2)$ family emerges from the Newtonian limit  of the  (red on Figure 1) $\OO(1)+\OO(1)$ curves . These are the
green curves on Figure 1.
\end{enumerate}
\begin{center}
\includegraphics[scale=0.4,angle=0]{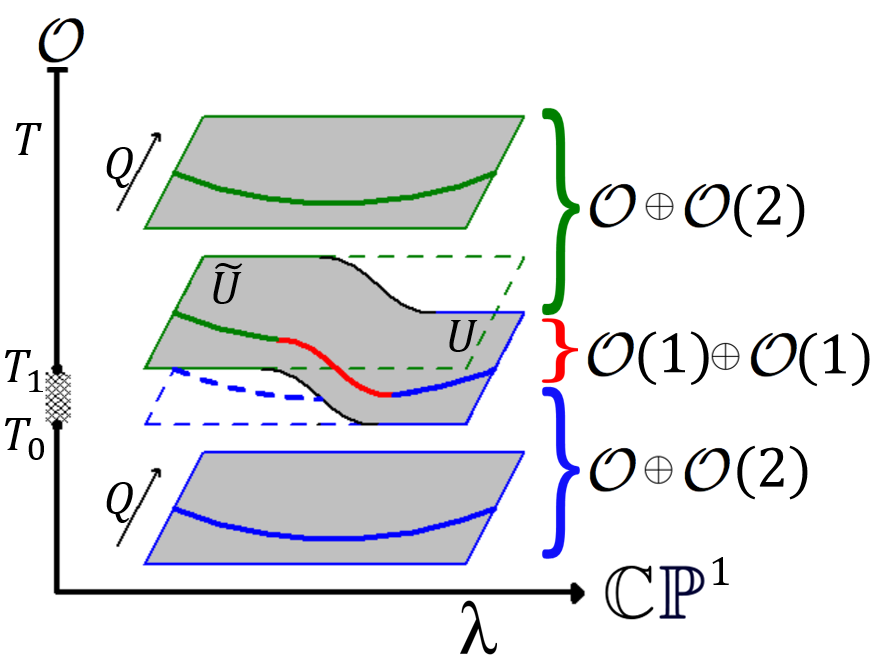}
\begin{center}
{{\bf Figure 1.} {\em Twistor collapse of the wave function}}
\end{center}
\end{center}

Thus, although the space time seems to bifurcate and collapse in the
{\bf R}-process, the twistor space is one complex three--fold. The curves in the
{\bf R}-process change their holomorphic type.


\begin{thebibliography}{jafsdl}
\frenchspacing 

\bibitem{ADM17} 
Atiyah, M., Dunajski, M. and Mason, L. (2017) Twistor theory at fifty: from contour integrals to twistor strings. 
{\tt arXiv:1704.07464}. Proceedings of the Royal Society, {\bf 473.}

\bibitem{CD} Cole, M and Dunajski, M. (2014) Twistor Theory of the Airy 
equation. {\tt arXiv:1401.0025}. SIGMA {\bf 10}, 037.

\bibitem{DG16} Dunajski, M. and Gundry, J. (2016)
Non-relativistic twistor theory and Newton--Cartan geometry. Comm. Math. Phys.
{\bf 342}, 1043-1074. {\tt arXiv:1502.03034.}


\bibitem{duval} Duval, C., Burdet, G.,  Kunzle, H and
Perrin, M.  (1985)  Bargmann structures and Newton-Cartan theory.
Phys. Rev. {\bf D31} 1841.

\bibitem{duval2} Duval, C., Gibbons, G. and Horv\'athy, P. (1991)  Celestial mechanics, conformal structures, and gravitational waves. Phys. Rev. {\bf D43}, 3907.


\bibitem{Hor}  
Duval, C.,  Horv\'athy, P. A. and Palla, L. (1994)
Conformal properties of Chern-Simons vortices in external fields, 
Phys. Rev. {\bf D50}, 6658.
  

\bibitem{Gundry} Gundry, J. (2017)
{\em Newtonian Twistor Theory}, PhD Thesis. University of Cambridge.

\bibitem{eisenhart} Eisenhart, L. P. (1928)
Dynamical trajectories and geodesics,
Annals. Math. {\bf 30} 591-606.


\bibitem{moroz} Moroz, I. Penrose, R. and Tod, K. P. (1998)
Spherically-symmetric solutions of the Schrödinger-Newton equations
Classical and Quantum Gravity, {\bf 15},

\bibitem{nish} Nesvizhevsky, V. V. et. al. (2002)
Quantum states of neutrons in the Earth's gravitational field.
Nature {\bf 415}, 297–299.


\bibitem{kodaira} Kodaira, K. (1963) On stability of compact submanifolds of complex manifolds, Am. J. Math. {\bf 85},
79-94.

\bibitem{Pen_2} Penrose, R. (1969), 
Solutions of the zero-rest-mass equations. J. Math. Phys. {\bf 10}, 38. 

\bibitem{Pen_3} Penrose, R. (1976) Nonlinear 
gravitons and curved twistor theory, Gen. Rel. Grav.  {\bf 7},  31--52

\bibitem{PR} Penrose, R. \& Rindler, W. (1986) {\em Spinors and
Space-Time}, Vol {\bf 1, 2}, CUP.

\bibitem{PRTN} Penrose, R.(1988)  Twistors and State-Vector Reduction. Twistor Newsletter {\bf 26.}

\bibitem{Pe96} Penrose, R.  (1996) 
{On gravity's role in quantum state reduction}, Gen. Rel.  Grav {\bf 28}.

\bibitem{Pe11} Penrose, R. (2014) 
On the gravitization of quantum mechanics 1: Quantum state reduction. 
Found. Phys. {\bf 44} 557-575.

\bibitem{P22} Penrose, R. (2022) New Physics for the Orch-OR Comciousness Proposal, in Consciousness and Quantium Mechanics (ed. Shan Gao)
Oxford University Press, Oxford UK.

\bibitem{tod_1} Tod, K. P. (1982) The singularities of H-space. Mathematical Proceedings of the Cambridge Philosophical Society, {\bf 92}, 331.


\end{thebibliography}
\end{document}